\documentclass[aps,pra,twocolumn,superscriptaddress,notitlepage,nofootinbib,longbibliography]{revtex4-2}
\usepackage{mathrsfs}
\usepackage{epsfig}
\usepackage{graphicx}
\usepackage{amsfonts}
\usepackage[figuresright]{rotating}
\usepackage{amssymb}
\usepackage{amsmath}
\usepackage{dcolumn}
\usepackage{bm}
\usepackage{color}
\usepackage{braket}
\usepackage{units}
\usepackage{xspace}

\usepackage{bbold}
\usepackage[colorlinks=true, allcolors=blue]{hyperref}
\definecolor{mypink3}{cmyk}{0, 0.7808, 0.4429, 0.1412}
\definecolor{mypink1}{rgb}{0.858, 0.188, 0.478}
\definecolor{mypink2}{RGB}{219, 48, 122}

\usepackage{float}
\begin{document}
\title{Photonic spin Hall effect in $\mathcal{PT}$-symmetric non-Hermitian cavity magnomechanics}
\author{Shah Fahad}
\affiliation{Department of Physics, Zhejiang Normal University, Jinhua, Zhejiang 321004, China}
\author{Muzamil Shah}
\affiliation{Department of Physics, Quaid-I-Azam University Islamabad, 45320, Pakistan}
\author{Gao Xianlong}
\email {gaoxl@zjnu.edu.cn}\affiliation{Department of Physics, Zhejiang Normal University, Jinhua, Zhejiang 321004, China}

%%%%%%%%%%%%%%%%%%%%%%%%%%%%%%%%%%%%%%%%%%%%%%%%%%%%%%%%%%%%%%%%%%%%%%%%%%%%%%%%%%%%%%%%%%%%%%%%%%%%%%%%%%%%%%%%%%%%%%%%%%%
\setlength{\parskip}{0pt}
\setlength{\belowcaptionskip}{-10pt}
\begin{abstract}
Non-Hermitian cavity magnomechanics, which incorporates the magnon-photon and magnon-phonon interactions simultaneously, enables rich physical phenomena, including exceptional-point-enhanced sensing, and offers pathways toward topological transitions and nonreciprocal quantum transformation. These interactions exert a pivotal influence on the optical response of a weak probe field and pave the way for novel applications in quantum technologies. In this work, we consider a yttrium-iron-garnet (YIG) sphere coupled to a microwave cavity. The magnon mode of the YIG sphere is directly excited through microwave field coupling, whereas the cavity mode is probed via a weak-field interrogation scheme. The direct interaction of a traveling field with the magnon mode induces gain in the system, thereby establishing non-Hermitian dynamics. The parity-time ($\mathcal{PT}$)-symmetric behavior of a hybrid non-Hermitian cavity magnomechanical system is designed and investigated. Eigenvalue spectrum analysis demonstrates that a third-order exceptional point ($\mathrm{EP}_3$) emerges under tunable effective magnon-photon coupling when the traveling field is oriented at an angle of $\pi/2$ relative to the cavity's $x$-axis. The photonic spin Hall effect (PSHE) in a reflected probe field is subsequently examined in such a system. Under balanced gain and loss conditions and in the presence of effective magnon-phonon coupling, tunable effective magnon-photon coupling enables coherent control of the PSHE across the broken $\mathcal{PT}$-symmetric phase, at the $\mathrm{EP}_3$, and in the $\mathcal{PT}$-symmetric phase. Investigation reveals that the PSHE can be significantly enhanced or suppressed via effective magnon-photon coupling. The influence of intracavity length on the PSHE is further explored, providing an additional parameter for fine-tuning the transverse shift. These findings establish a direct correspondence between the PSHE and the underlying non-Hermitian eigenvalue spectrum. They further demonstrate that non-Hermitian cavity magnomechanical platforms provide a pathway to tunable photonic functionalities, with immediate prospects for quantum switching and precision sensing.
\end{abstract}
%\pacs{42.50.Pq, 42.50.Gy, 67.85.Hj, 71.70.Ej}
\date{\today}
\maketitle
\section{Introduction}
Magnons, the collective spin excitations in magnetic materials~\cite{Serga2010}, have attracted increasing attention as a platform for examining macroscopic quantum phenomena~\cite{LENK2011, Liu2019, CUi2019}. In particular, the cavity magnomechanics system combines magnons, photons, and phonons into a single physical platform, bridging research fields that span quantum information, cavity quantum electrodynamics, magnonics, and quantum optics~\cite{zuo2024}. This unified framework enables precise characterization of both semi-classical and quantum behaviors, leading to such notable effects as magnon bistability~\cite{Wang2016Magnon, Hyde2018Direct}, cavity-magnon polaritons~\cite{Cao2015Exchange, Yao2015Theory}, high-order sideband generation~\cite{xu2020magon}, and the creation of entanglement and squeezed states~\cite{YU2020, Qiu2022, Li2019sqeez, Li2022}. Beyond these, cavity magnomechanics offers routes to controllable magnonic switching~\cite{Hao2024controllabe}, magnon dark modes~\cite{Zhang2015Magnon}, and exotic non-Hermitian effects~\cite{Harder2017}.

Through the coherent coupling of magnons, phonons, and cavity microwave photons, the microwave response of such systems features magnon-induced absorption and magnomechanically induced transparency~\cite{Wen2019, Wang2018, Kamran2020, Li2020phasecontrol, Lu2021Ep, Munir2023}. These phenomena can be understood via interference mechanisms analogous to those in cavity optomechanics, highlighting the interplay of constructive and destructive pathways within the system~\cite{Hou2015, Agrawal2010, Zhang2017}. Additionally, cavity magnomechanics has enabled demonstrations of tunable slow and fast light~\cite{Liu2019Slowfast, Wen2019, CUi2019} and opened new avenues for combining parity-time ($\mathcal{PT}$)-symmetric paradigms with higher-order sideband control~\cite{Huai2019Enhanced}. 

Unavoidable coupling to the environment in realistic systems induces decoherence~\cite{Weiss2012} and is effectively captured by a non-Hermitian Hamiltonian $H \neq H^\dagger$~\cite{Ozdemir2019, Minganti2019}. For pseudo-Hermitian Hamiltonians—those satisfying $U H U^{-1} = H^\dagger$ with a linear Hermitian operator $U$~\cite{MostafazadehI, MostafazadehII}—eigenvalues are either real or occur in complex-conjugate pairs. A prominent subclass is provided by 
$\mathcal{PT}$-symmetric Hamiltonians, defined by $[H, \mathcal{PT}] = 0$~\cite{KonotopRMP, HOEPPRA2021, Zhang2025}, which embody a characteristic balance of gain and loss. Varying a single control parameter generically drives a $\mathcal{PT}$-symmetric Hamiltonian through a second-order exceptional point ($\mathrm{EP}_2$), where two eigenvalues and their eigenvectors coalesce, marking the transition from the $\mathcal{PT}$-symmetric phase (real spectrum) to the $\mathcal{PT}$-broken phase (complex-conjugate spectrum)\cite{Bender2013, Liu2017, Midya2021}. $\mathrm{EP}_2$ physics has been extensively explored across platforms including cavity optomechanics~\cite{Liu2017, JingPRL, Xu2016}, waveguides~\cite{Doppler2016}, microcavities~\cite{Chang2014}, cavity magnonics~\cite{Dengke2017, Harder2017}, and cavity magnomechanics~\cite{Lu2021Ep}. Beyond $\mathrm{EP}_2$, non-Hermitian systems can host higher-order exceptional points, at which more than two eigenmodes coalesce~\cite{Heiss2016, HOEPPRA2021, ZhangPRB2019, Demange2011, Heiss2008}. Such higher-order singularities exhibit stronger non-Hermitian degeneracies than $\mathrm{EP}_2$s and enable enhanced sensing~\cite{Hodaei2017, Zeng2021}, richer topological responses~\cite{DingPRX, DelplacePRL}, and pronounced spontaneous-emission enhancement~\cite{LinPRL}.

Beyond these explorations, notable applications arise in controlling the transmission of output microwave fields via $\mathcal{PT}$-symmetric interactions~\cite{Das2023, Wang2023}. Underlying many of these phenomena is the concept of magnomechanically induced transparency~\cite{Kamran2020, Bayati2024}, whereby strong magnon-photon interaction creates a transparent window in the absorption spectrum. Furthermore, the presence of magnon-phonon coupling can engineer multiple transmission windows by suitably adjusting the phase and amplitude of an external magnetic field~\cite{Li2020phasecontrol}. Intriguingly, the resulting tunable optical response, along with the attendant group delay control, has been systematically investigated~\cite{Wahab2025}, underscoring how fundamental non-Hermitian principles open possibilities for applications in quantum metrology and device engineering. 

In this work, we investigate the photonic spin Hall effect (PSHE) as a key aspect of the optical response in a non-Hermitian cavity magnomechanical system. This hybrid platform offers new insights into spin-dependent light-matter interactions and potential control mechanisms for spin-polarized photonic transport.

The PSHE manifests as a spin-dependent spatial separation of photons via spin-orbit coupling~\cite{Onoda2024,kim2023spin}, inducing a transverse shift in the incident plane for linearly polarized beams composed of right- and left-circular polarization components. This phenomenon constitutes a photonic analog of the electronic spin Hall effect, where refractive index gradients and photon spin mirror the roles of electric potentials and electron spin~\cite{Bliokh2006,hosten2008}. Historically attributed to the Imbert-Fedorov displacement~\cite{Imbert1972, Fedorov2013}, the transverse shift has been extensively studied in diverse systems including metamaterials~\cite{yin2013photonic}, topological insulators~\cite{SHAH2022surface}, graphene heterostructures~\cite{zhou2012identifying, SHAH2024tunable, Cai2017}, 2D quantum materials~\cite{Kort2017, Shah2022}, plasmonic resonators~\cite{Luca2012, Zhou2016Enhaced, Tan2016}, gravitational fields~\cite{Gosselin2007}, and semiconductors~\cite{Menard09}.

While the PSHE has been examined in conventional optical and cavity magnomechanical settings~\cite{MuqddarPRA, MUNIR2025}, its behavior in 
$\mathcal{PT}$-symmetric non-Hermitian cavity magnomechanical systems—across the unbroken and broken $\mathcal{PT}$ phases and at the third-order exceptional point 
$\mathrm{EP}_3$—has not been systematically characterized. Here we address this gap by analyzing a hybrid three-mode platform with magnon–photon and magnon–phonon couplings. We show that, under experimentally feasible parameters, these interactions control spin-dependent transverse shifts: the PSHE exhibits phase-selective responses and pronounced features at $\mathrm{EP}_3$, with its magnitude enhanced or suppressed depending on the underlying non-Hermitian eigenvalue structure. Our results elucidate how $\mathcal{PT}$ symmetry and higher-order exceptional points govern the PSHE in cavity magnomechanics, providing routes to tunable photonic functionalities in engineered non-Hermitian systems.

In this paper, we investigate a non-Hermitian cavity magnomechanical system across broken-$\mathcal{PT}$, $\mathrm{EP}_3$, and $\mathcal{PT}$-symmetric phases. The system comprises a microwave cavity with two fixed mirrors and an embedded yttrium-iron-garnet (YIG) sphere. A uniform $B_z$-field along the $z$-direction induces magnon modes in the YIG sphere, coupled to cavity photons via magnetic dipole interaction. Magnetostrictive forces link magnons to phonons via lattice deformation. Non-Hermiticity is engineered through a traveling field driving the magnon mode, introducing gain. At gain–loss balance, we demonstrate a third-order $\mathrm{EP}_{3}$ by tuning the effective magnon-photon coupling and traveling-field incidence angle.
Using the transfer matrix method, we analyze PSHE in reflected probe fields, demonstrating control across all $\mathcal{PT}$ phases. Transverse shifts derived from Fresnel coefficients reveal enhanced PSHE in the $\mathcal{PT}$-symmetric phase relative to the broken phase, while $\mathrm{EP}_3$ exhibits suppression—consistent with PSHE control via exceptional points~\cite{controllingPSHE}. PSHE modulation via intracavity length variation is also demonstrated. These results establish a dynamic control mechanism for spin-dependent photonic transport in non-Hermitian platforms.

The rest of the manuscript is structured as follows: Section II details the physical system and Hamiltonian, Sec. III presents results and discussion, and Sec. V concludes.

\section{System Hamiltonian}
\begin{figure}
\includegraphics[width=\linewidth]{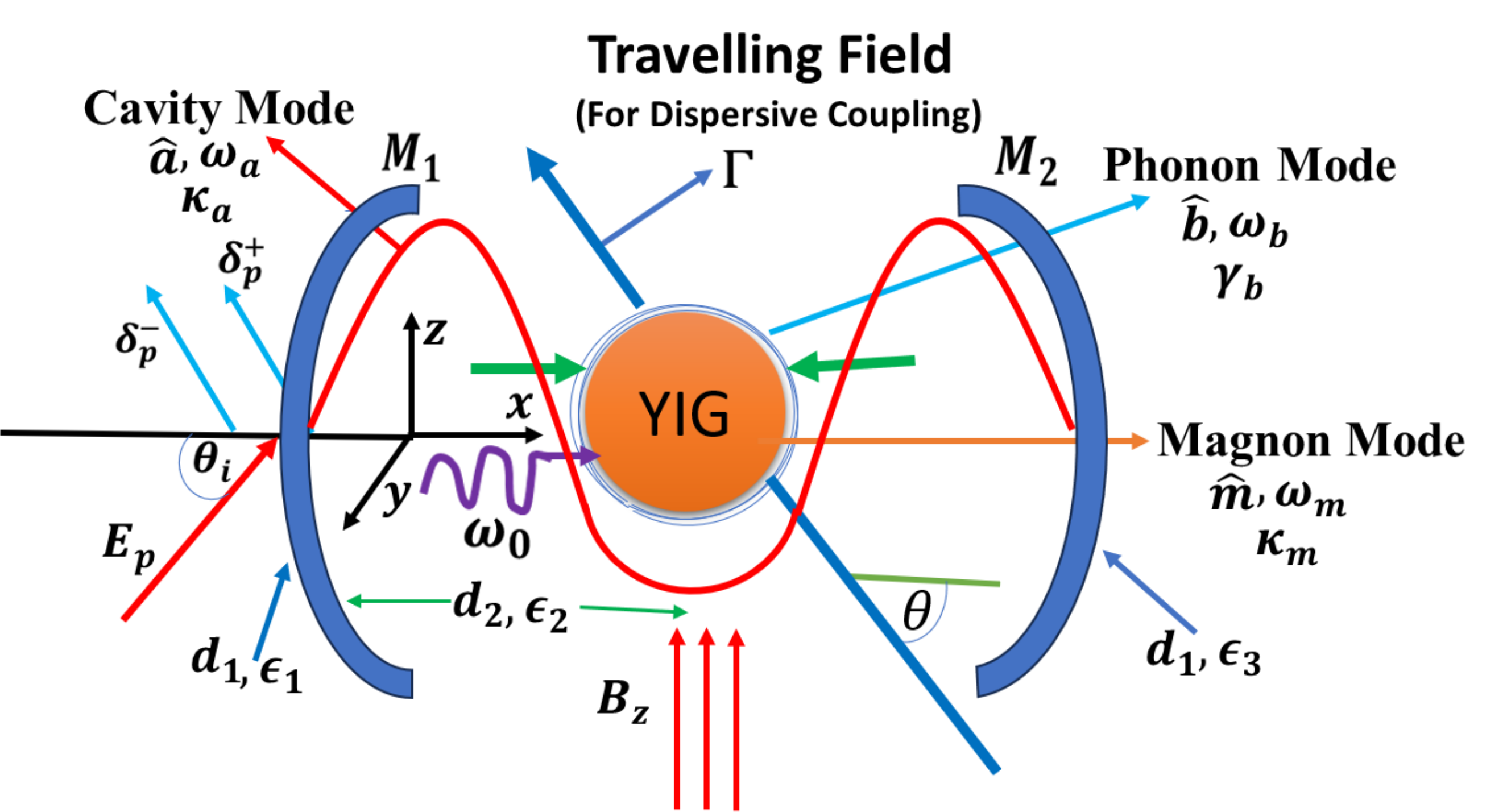}
\caption{Schematic illustration of a non-Hermitian cavity magnomechanical system with an embedded YIG sphere. A microwave cavity (photon mode $\hat{a}$, resonance frequency $\omega_a$, dissipation $\kappa_a$) hosts the sphere under $z$-axis bias field $B_z$ (exciting magnon mode $\hat{m}$, resonance frequency $\omega_m$, gain $\kappa_m$). Magnon-photon coupling occurs via magnetic dipole interaction. Magnetostriction excites phonon mode $\hat{b}$ ($\omega_b$, dissipation $\gamma_b$), enabling magnon-phonon coupling enhanced by $x$-axis microwave drive field ($\omega_0$). Orthogonal cavity ($B_y$), bias ($B_z$), and drive ($B_x$) magnetic fields are shown. Non-Hermiticity arises from a traveling field (incident angle $\theta$, coupling $\Gamma$). A TM-polarized probe field $E_p$ incident on mirror $M_1$ at $\theta_i$ undergoes spin-dependent splitting; reflected transverse shifts $\delta^{\pm}_p$ for left/right circular components are measured.}
\label{fig1}
\end{figure}

We consider a non-Hermitian cavity magnomechanical platform comprising a single-mode microwave cavity $\hat{a}$ (resonance frequency $\omega_{a}$) and a single-crystal YIG sphere (Fig.~\ref{fig1}). The YIG sphere is positioned at the magnetic-field antinode of the cavity mode along the $x$-axis. The cavity is implemented as a three-layer structure formed by two nonmagnetic mirrors $M_1$ and $M_2$ separated by a fixed distance $d_2$; $M_2$ is perfectly reflecting, whereas $M_1$ is partially reflecting. Both mirrors have thickness $d_1$ and permittivities $\epsilon_1$ and $\epsilon_3$, and the intracavity medium has effective permittivity $\epsilon_2$, in direct analogy with cavity optomechanical~\cite{Muhib2019} and atomic systems~\cite{Wang2008}. A uniform $z$-axis bias field $B_z$ excites a magnon mode $\hat{m}$ (resonance frequency $\omega_{m}$) in the YIG sphere, which couples to the cavity photons $\hat{a}$ via the magnetic-dipole interaction. Owing to its material quality and geometry, the YIG sphere also functions as a high-$Q$ mechanical resonator $\hat{b}$ (frequency $\omega_b$). Magnetization dynamics associated with $\hat{m}$ drive lattice deformation via magnetostriction, thereby mediating a magnon-phonon coupling through the magnetostrictive force. The resulting hybrid three-mode system supports photon-magnon and magnon-phonon interactions and provides a versatile setting for exploring non-Hermitian cavity magnomechanical phenomena.
 
The system Hamiltonian is given by
\begin{equation}
\hat{\mathcal{H}}=\hat{\mathcal{H}}_{0}+\hat{\mathcal{H}}_{c}+\hat{\mathcal{H}}_{d}+\hat{\mathcal{H}}_{p}+\hat{\mathcal{H}}_{non},\label{Total Hamiltonian}
\end{equation}
where $\hat{\mathcal{H}}_{0} = \hbar\omega _{a}\hat{a}^{\dagger}\hat{a}+  \hbar\omega _{m}\hat{m}^{\dagger}\hat{m}+\hbar\omega_{b}\hat{b}^{\dagger}\hat{b}$ describes uncoupled mode energies, $\hat{\mathcal{H} }_{c} = \hbar g_{ma}(\hat{a}+ \hat{a}^{\dagger })(\hat{m}+\hat{m}^{\dagger}) + \hbar g_{mb}\hat{m}^{\dagger}\hat{m}(\hat{b}^{\dagger}+\hat{b})$ governs coherent couplings, $\hat{\mathcal{H}}_{d} = i \hbar \eta(\hat{m}^{\dagger}\mathrm{e}^{-i\omega_{0}t}-{\rm H.c.})$ drives the magnon mode, $\hat{\mathcal{H}}_{p}=i \hbar E_{p}(\hat{a}^{\dagger}\mathrm{e}^{-i\omega_{p}t}-{\rm H.c.})$ couples the probe field, and $\hat{\mathcal{H}}_{non} = -i\hbar \Gamma \mathrm{e}^{i(\omega t+\theta) }(\hat{a}+\hat{a}^{\dagger })(\hat{m}+\hat{m}^{\dagger })$ introduces non-Hermiticity.
Here, $\omega_{a,m,b}$ denote photon, magnon, and phonon frequencies; $\hat{a}$, $\hat{b}$, $\hat{m}$ ($\hat{a}^\dagger$, $\hat{b}^\dagger$, $\hat{m}^\dagger$) are annihilation (creation) operators; $g_{ma}$ ($g_{mb}$) is the magnon-photon (magnon-phonon) coupling. The magnon mode is driven by a microwave field (frequency $\omega_0$, amplitude $\eta= \sqrt{5N}\gamma B_{0}/4$, with the total number of spins $N$, 
%YIG volume $V$, 
the external magnetic field $B_{0}$, gyromagnetic ratio $\gamma$). A weak probe field (frequency $\omega_p$, power $P$, incident angle $\theta_i$) couples to the cavity via $E_{p} = \sqrt{2P\kappa_{a}/\hbar\omega_{p}}$. The non-Hermitian term $\hat{\mathcal{H}}_{non}$ arises from a traveling field (coupling $\Gamma=\frac{\omega_{a}}{d_{2}}\sqrt{(\hbar/\omega_{m} m_{m})}$ with the magnon mass $m_{m}$)~\cite{pramanik2020,GePRA2013,Teklu2018}). 

Employing the rotating wave approximation~\cite{Li2018Magnon, Wenlin2016Parity}, which neglects rapidly oscillating terms under weak-coupling conditions, we transition to a rotating frame where the traveling field becomes effectively time-independent. This field couples to the YIG sphere and introduces gain for the magnon mode. Consequently, Eq.~(\ref{Total Hamiltonian}) reduces to:
\begin{align}
\hat{\mathcal{H}} &= \hbar\Delta_{a}\hat{a}^{\dagger}\hat{a}+ \hbar\Delta_{m}\hat{m}^{\dagger}\hat{m}+ \hbar\omega_b\hat{b}^{\dagger}\hat{b} \nonumber\\ 
& +\hbar(g_{ma}-i \Gamma \mathrm{e}^{i \theta})(\hat{a}\hat{m}^{\dagger}+\hat{a}^{\dagger}\hat{m}) + \hbar g_{mb}\hat{m}^{\dagger}\hat{m}(\hat{b}^{\dagger}+\hat{b}) \nonumber\\ 
&+ i \hbar\eta(\hat{m}^{\dagger}-\hat{m}) + i\hbar E_{p}(\hat{a}^{\dagger}\mathrm{e}^{-i \Delta_{p} t}-\hat{a}\mathrm{e}^{i \Delta_{p} t}).\label{simplified-H}
\end{align}
Here $\Delta_{a} = \omega_{a} - \omega_{0}, \quad \Delta_{m} = \omega_{m} - \omega_{0},\ \text{and} \quad \Delta_{p}=\omega_{p}-\omega_{0}$.

The dynamics follow from the Heisenberg-Langevin equations for operators $\hat{O} \in {\hat{a}, \hat{m}, \hat{b}}$:
\begin{equation}
\frac{d\hat{O}}{dt} =  \frac{i}{\hbar} {[\hat{\mathcal{H}}, \hat{O}]} -\zeta \hat{O},\label{Generic form}  
\end{equation}
where $\zeta$ denotes the decay ($\zeta>0$) or gain ($\zeta<0$) rate, and $[\hat{O}, \hat{O}^\dagger] = 1$ for $\hat{O} \in {\hat{a}, \hat{m}, \hat{b}}$. Neglecting quantum and thermal noise~\cite{waseemGoos2024}, we obtain:
\begin{align}
\dot{\hat{a}}&= -(i\Delta_{a}+\kappa_{a})\hat{a}-(ig_{ma}+\Gamma \mathrm{e}^{{i}\theta })\hat{m} + E_{p}\mathrm{e}^{-i\Delta_{p} t},\nonumber\\    
\dot{\hat{m}}&= -(i\Delta_{m} - \kappa_{m})\hat{m} -(ig_{ma}+\Gamma \mathrm{e}^{i\theta})\hat{a}-ig_{mb}\hat{m}(\hat{b}^\dagger + \hat{b}) + \eta,\nonumber\\ 
\dot{\hat{b}}&= -(i\omega_{b}+\gamma_{b})\hat{b} -ig_{mb}\hat{m}^{\dagger}\hat{m}.\label{HLEs-1} 
\end{align}
Here $\kappa_a$ ($\gamma_b$) is the photon (phonon) dissipation rate, and $\kappa_m$ the magnon gain. Replacing operators with expectations $O(t) \equiv \langle \hat{O}(t) \rangle$ ($O = a, m, b$)~\cite{Xiong2015} yields:
 \begin{align}
\dot{a}&= -(i\Delta_{a}+\kappa_{a})a-(ig_{ma}+\Gamma \mathrm{e}^{{i}\theta })m + E_{p}\mathrm{e}^{-i\Delta_{p} t},\nonumber\\ 
\dot{m}&= -(i\Delta_{m} - \kappa_{m})m -(ig_{ma}+\Gamma \mathrm{e}^{i\theta})a-ig_{mb}m(b^\ast + b) + \eta,\nonumber\\ 
\dot{b}&= -(i\omega_{b}+\gamma_{b})b -ig_{mb}m^{\ast}m.\label{HLEs-2} 
\end{align}
Steady-state solutions ($O_s$) satisfy:
\begin{align}
a_{s}&= \frac{-(ig_{ma} +\Gamma {e}^{i\theta})m_{s}}{( i\Delta_{a}+\kappa_{a})},\nonumber\\
m_{s}& = \frac{-(ig_{ma}+\Gamma \mathrm{e}^{i\theta}){{a}_{s}} +\eta}{(i\Delta_{s} - \kappa_{m})},\nonumber\\
b_{s} &= \frac{-ig_{mb} |{m_{s}}|^2}{(i\omega_{b}+\gamma_{b})},\label{steady states}
\end{align}
where $\Delta_{s}= \Delta_{m} + g_{mb}({b}^{\ast}_{s} +{b_s})$ is the effective magnon-phonon detuning.
Linearizing via $O = O_s + \delta O$ and retaining first-order terms gives:
 \begin{align}
\delta\dot{a}& = -(i\Delta_{a} + \kappa_{a})\delta{a}-(ig_{ma} + \Gamma {e}^{i\theta})\delta{m} + E_{p}\mathrm{e}^{-i \Delta_{p} t},\nonumber\\
\delta\dot{m}&= - (i\Delta_{s} - \kappa_{m})\delta{m}-(ig_{ma} +\Gamma \mathrm{e}^{i\theta})\delta {a}-ig_{mb}m_{s}\delta{b},\nonumber\\ 
\delta\dot{b}& =- (i\omega_{b} + \gamma_{b}) \delta{b}-ig_{mb}m^{\ast}_{s}\delta{m}. \label{linerized-1}  
 \end{align}

\subsection{Effective Hamiltonian and $\mathcal{PT}$-symmetry}\label{SecIIA}

\begin{figure}
\centering
\includegraphics[width=\linewidth]{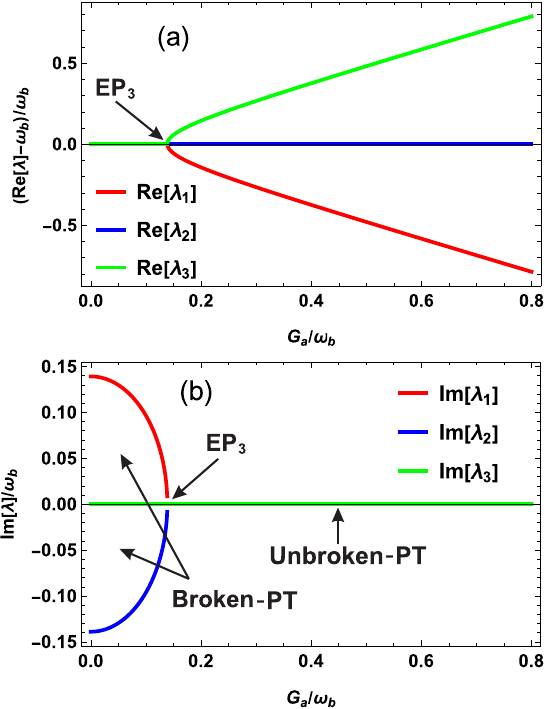}
\caption{Eigenvalues of $H_{\text{eff}}$ [Eq.~(\ref{H_eff})] versus normalized effective magnon-photon coupling $G_{a}/\omega_{b}$. (a) Real part: $(\operatorname{Re}[\lambda]-\omega_{b})/\omega_{b}$; (b) imaginary part: $\operatorname{Im}[\lambda]/\omega_{b}$. Parameters: $\kappa_{a}/2\pi= 2.1~\mathrm{MHz}$, $\gamma_{b}/2\pi= 150~\mathrm{Hz}$, $\kappa_{m}=\kappa_{a} + \gamma_{b}$, $\omega_{b}/2\pi =15.101~\mathrm{MHz}$, $\Delta_{s}/2\pi=\Delta_{a}/2\pi=15.10~\mathrm{MHz}$, and $G_{b}/2\pi=0.001~\mathrm{MHz}$.}
\label{fig-2}
\end{figure}
The Heisenberg–Langevin equations in Eq.~(\ref{linerized-1}) can be expressed in matrix form as:
\begin{equation}
 \dot{\mathbf{O}} = -i H_{\text{eff}}\,\mathbf{O},
\end{equation}
where $\mathbf{O}= (\delta a, \delta m, \delta b)^\mathsf{T}$ is the column vector, and $H_{\text{eff}}$ represents the effective non-Hermitian Hamiltonian of the hybrid system,
\begin{equation}
H_{\text{eff}}=
{\begin{pmatrix}
\Delta_{a} -i\kappa_{a} & g_{ma} -i\Gamma\mathrm{e}^{i\theta} & 0 \\
g_{ma} -\mathrm{i\Gamma}\mathrm{e}^{i\theta}  & \Delta_{s} + i\kappa_{m}  &  G_{b} \\ 0 & G^*_{b} & \omega_{b} - i\gamma_{b} 
\end{pmatrix}}.\label{H_eff}
\end{equation}
Here, $G_{b} = g_{mb}m_{s}$ is the effective magnomechanical (magnon-phonon) coupling coefficient. This hybrid three-mode system with magnon-photon and magnon–phonon couplings yields three eigenvalues. The eigenvalues $\lambda$ are determined by:
 
\begin{equation}
\lambda^3 + r\lambda^2 + s\lambda + t=0,\label{3rd Eq}
\end{equation}
where $r=-[\Delta_{a} -i\kappa_{a} +\Delta_{s} +i\kappa_{m} +\omega_{b} -i\gamma_{b}]$, $s=(\Delta_{a} -i\kappa_{a})(\Delta_{s} +i\kappa_{m} +\omega_{b} -i\gamma_{b}) + (\Delta_{s} +i\kappa_{m})(\omega_{b} -i\gamma_{b})-|G_{b}|^2 - (g_{ma} -i\Gamma\mathrm{e}^{i\theta})^2$, $t=-(\Delta_{a}-i\kappa_{a})(\Delta_{s} +i\kappa_{m})(\omega_{b} -i\gamma_{b}) + |G_{b}|^2(\Delta_{a} -i\kappa_{a}) +(\omega_{b} -i\gamma_{b})(g_{ma} -i\Gamma\mathrm{e}^{i\theta})^2$.
The magnon mode is assumed to be driven by a microwave field in the red-sideband regime $(\Delta_{a}= \Delta_{s}\approx\omega_{b})$~\cite{waseemGoos2024, He2024}. We establish pseudo-Hermiticity conditions for the effective Hamiltonian $H_{\text{eff}}$ [Eq.~(\ref{H_eff})]. Following Refs.~\cite{HOEPPRA2021, Zhang2025}, $H_{\text{eff}}$ is pseudo-Hermitian if: (i) all eigenvalues are real, or (ii) one is real and the others form a complex-conjugate pair. At $\theta=\pi/2$, the spectrum satisfies $\text{det}(H_{\text{eff}} - \lambda I)= \text{det}(H_{\text{eff}}^\ast - \lambda I)= 0$, where $I$ is the identity matrix. The parameters of the system should satisfy the following constraints: 
\begin{equation}
\begin{aligned} 
\kappa_{a} + \gamma_{b} - \kappa_{m}  =0,\\
\kappa_{a}\kappa_{m}\gamma_{b} - \kappa_{a}|G_{b}|^2 - \gamma_{b}(g_{ma} -i\Gamma\mathrm{e}^{i\theta})^2=0.\label{conditions} 
\end{aligned} 
\end{equation}
\begin{figure*}
\begin{center}
\includegraphics[width=\linewidth]{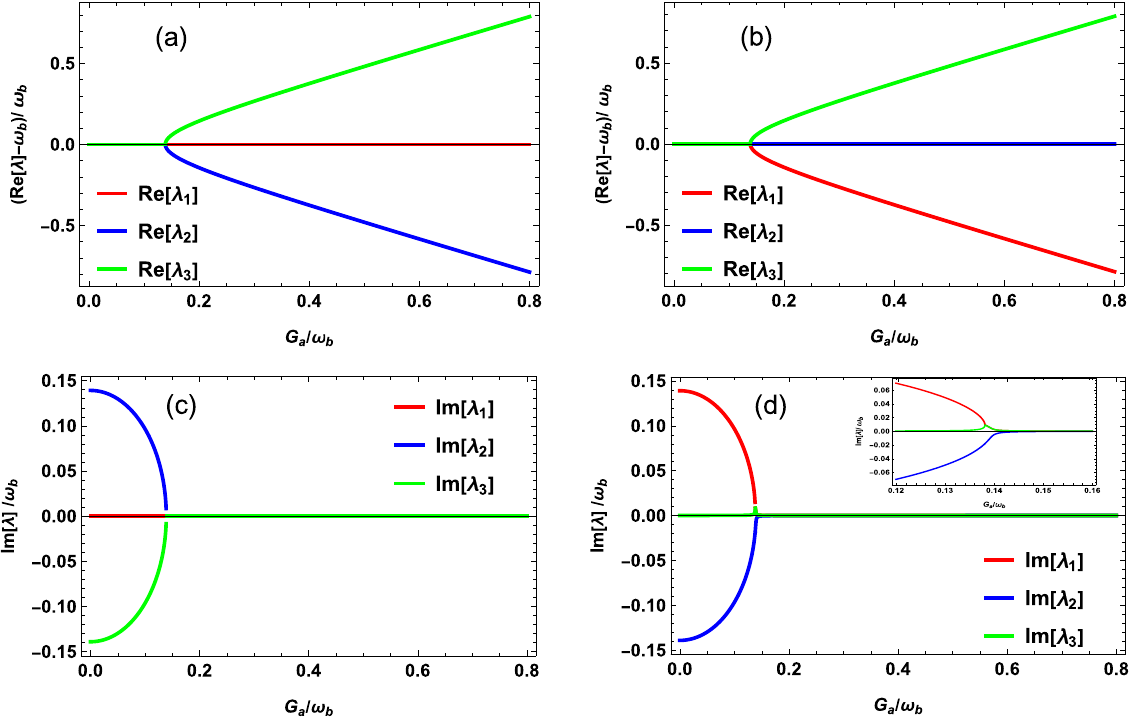}
\caption{Eigenvalues of $H_{\text{eff}}$ [Eq.~(\ref{H_eff})] as a function of $G_{a}/\omega_{b}$. (a,b) Real parts: $(\operatorname{Re}[\lambda]-\omega_{b})/\omega_{b}$; (c,d) Imaginary parts: $\operatorname{Im}[\lambda]/\omega_{b}$. Columns correspond to effective magnon–phonon coupling strengths $G_{b}=0$ and $G_{b}/2\pi=0.05~\mathrm{MHz}$, respectively. The inset in Fig.~\ref{fig2b}(d) provides a magnified view of $\operatorname{Im}[\lambda]/\omega_{b}$ as a function of $G_{a}/\omega_{b}$. Fixed parameters: $\kappa_{a}/2\pi = 2.1~\mathrm{MHz}$, $\gamma_{b}/2\pi = 150~\mathrm{Hz}$, $\kappa_{m} = \kappa_{a} + \gamma_{b}$, $\omega_{b}/2\pi = 15.101~\mathrm{MHz}$, $\Delta_{s}/2\pi =\Delta_{a}/2\pi = 15.10~\mathrm{MHz}$, and at $G_{b}=0$, $\Delta_{m}/2\pi = \Delta_{a}/2\pi = 15.10~\mathrm{MHz}$.}
\label{fig2b}
\end{center}
\end{figure*}
According to Eq.~\eqref{conditions}, the effective Hamiltonian $H_{\text{eff}}$ is pseudo-Hermitian. Under the balanced gain-loss condition ($\kappa_m = \kappa_a + \gamma_b$) and a traveling-field phase $\theta = \pi/2$, $H_{\text{eff}}$ exhibits a $\mathcal{PT}$-symmetric eigenvalue spectrum for  $0 < G_b \lesssim  2\pi \times 0.01~\mathrm{MHz}$, a parameter window contained within the broader pseudo-Hermitian regime~\cite{HOEPPRA2021, KonotopRMP, Zhang2025}. Henceforth, we classify and discuss the system’s behavior in terms of the corresponding $\mathcal{PT}$ phases.

Figure~\ref{fig-2} (a) and (b) display the eigenvalue spectrum of the non-Hermitian cavity magnomechanical system versus normalized effective magnon-photon coupling $G_a / \omega_b$, where $G_{a}=(g_{ma} -i\Gamma\mathrm{e}^{i\theta})$ at $\theta=\pi / 2$. Under balanced gain-loss ($\kappa_m=\kappa_a+\gamma_b$), the broken $\mathcal{PT}$-symmetric phase ($G_a / \omega_b<0.139$) exhibits complex eigenvalues with nonzero real and imaginary components. At $G_a / \omega_b=0.139$, eigenvalue coalescence forms a third-order exceptional point ($\mathrm{EP}_3$). For $G_a / \omega_b>0.139$, the unbroken $\mathcal{PT}$-symmetric phase features three real eigenvalues.

To assess the impact of the effective magnon-phonon coupling strength $G_b$ on the spectrum, we analyze two representative cases: $G_b=0$ and $G_b=2\pi\times 0.05~\mathrm{MHz}$. For $G_b=0$, Eq.~(\ref{3rd Eq}) reduces to $(\lambda - (\omega_b - i\gamma_b))(\lambda^2 + A\lambda + B) = 0$, where $A = -(2\omega_b + i\gamma_b)$ and $B = \omega_b^2 + i\omega_b\gamma_b + \kappa_a^2 + \kappa_a\gamma_b - G_a^2$. In this uncoupled case, the mechanical mode eigenvalue $\lambda_1 = \omega_b - i\gamma_b$ remains distinct from the magnon-photon subsystem eigenvalues $\lambda_{2,3}$, and coalescence occurs only within the magnon–photon subsystem. The corresponding eigenvalues of $H_{\text{eff}}$ as functions of $G_a/\omega_b$ are shown in Fig.~\ref{fig2b}. For $G_b=0$, an  $\mathrm{EP}_2$ is clearly visible [panel~(a)]. Although the Hamiltonian remains formally $3\times 3$ due to the phonon term ($\omega_b - i\gamma_b$), the phonon mode is decoupled from the magnon-photon subsystem. The characteristic polynomial is therefore cubic, but a genuine  $\mathrm{EP}_3$ cannot arise because the three modes are not mutually coupled. The apparent triple root in [panel~(a)] occurs when $\omega_b - i\gamma_b$ coincides with a doubly degenerate eigenvalue of the magnon-photon sector; this is an algebraic (accidental) degeneracy without coalescence of three eigenvectors, and thus not a robust  $\mathrm{EP}_3$. In this limit, the system effectively reduces to a non-Hermitian magnonic model rather than a magnomechanical one. For  $G_b=2\pi\times 0.05~\mathrm{MHz}$, the  $\mathrm{EP}_3$ is lifted [panel~(b)], highlighting $G_b$ as a key control parameter of the spectral structure. In the experimentally relevant regime where $G_b$ is on the order of kHz—much smaller than the remaining (MHz-scale) parameters [Eq.~(\ref{3rd Eq})]—the spectrum remains $\mathcal{PT}$-symmetric for $0 < G_b \lesssim 2\pi\times 0.01~\mathrm{MHz}$, and the $\mathrm{EP}_3$ persists [Fig.~\ref{fig-2}]. Increasing $G_b$ beyond this window lifts the triple degeneracy and eliminates the  $\mathrm{EP}_3$ [panel~(b)].

\subsection{Optical Susceptibility}

To find the optical susceptibility of the coupled system, we solve Eq.~(\ref{linerized-1}) by introducing the slowly varying operator for the linear terms of the fluctuation as $\delta a \rightarrow \delta a\mathrm{e}^{-i\Delta_{a}t}$, $\delta m \rightarrow \delta m\mathrm{e}^{-i\Delta_{s}t}$, and $\delta b \rightarrow \delta b\mathrm{e}^{-i\omega_{b}t}$. Then Eq.~(\ref{linerized-1}) can be written as
\begin{equation}
 \begin{aligned}
\delta\dot{a}& = -\kappa_{a}\delta a-(ig_{ma} + \Gamma \mathrm{e}^{i\theta}) \delta m + E_{p}\mathrm{e}^{-ixt},\\
\delta\dot{m}&= + \kappa_{m}\delta {m}-(ig_{ma} +\Gamma \mathrm{e}^{i\theta})\delta{a} -ig_{mb}m_{s}\delta b,\\ 
\delta\dot{b}& = - \gamma_{b} \delta{b} -ig_{mb}m^{*}_{s}\delta m, \label{linerized-2} 
 \end{aligned}
\end{equation}
where $x=\Delta_{p}-\omega_{b}$ is the effective detuning. To solve Eq.~(\ref{linerized-2}), we apply the ansatz $\delta A = \delta{A}_{1}\mathrm{e}^{-ixt} + \delta A_{2}\mathrm{e}^{ixt}$ with $A= (a, m, b)$. Therefore, we obtain the first-order sideband amplitude $\delta{a}_{1}$ of the non-Hermitian cavity magnomechanical system for a weak probe field:
\begin{equation}
\delta{a}_{1} = \frac{E_{p}}{(\kappa_{a}-ix) + \frac{(\gamma_{b}-ix)(ig_{ma} + \Gamma\mathrm{e}^{i\theta})^2}{(\kappa_{m}+ix)(\gamma_{b} -ix) - G^2_{b}}}.\label{amplitude} 
\end{equation}
In this analysis, the contribution from $\delta{a_{2}}$ is negligible, as it corresponds to a four-wave mixing process at frequency 
$\omega_{p}-2\omega_{0}$ induced by the interaction between the weak probe field and the driving field. The output field 
$E_{T}$ of the weak probe field defines the optical susceptibility 
 $\chi$  through the relation~\cite{waseemGoos2024,MUNIR2025,Chen2023,Li2016Transparency}
\begin{equation}
\chi \equiv E_T = \kappa_{a} \delta a_{1} / E_{p},
\end{equation}
where 
$\chi$ is a complex quantity expressed in quadrature components as 
$\chi=\chi_{r} +i\chi_{i}$. These quadratures are measured via homodyne detection~\cite{walls1994quantum}. The real part 
$\chi_{r}$ governs the absorption spectrum, while the imaginary part 
$\chi_{i}$ determines the dispersion spectrum of the weak probe field.

The PSHE constitutes an optical analog of the electronic SHE~\cite{Bliokh2015}. The reflection coefficients for TE polarization ($r^{s}$) and TM polarization ($r^{p}$) are computed for the cavity using the transfer matrix method. The transfer matrix of the effective three-layer system is given by
\begin{widetext}
\begin{equation}
M^{s,p}_{j}(\theta_{i},\omega_{p},d_{j})=
{\begin{pmatrix}
\cos[{n_{j}k\cos{(\theta_{j})}d_{j}]}  & i\beta^{s,p}_{j}\sin[{n_{j}k\cos{(\theta_{j})}d_{j}]}/[n_{j}\cos{(\theta_{j})}] \\
i\,n_{j}\cos(\theta_{j})\sin{[{n_{j}k\cos{(\theta_{j})}d_{j}]}}/\beta^{s,p}_{j} & \cos[{n_{j}k\cos{(\theta_{j})}d_{j}]}\label{transfer matrix}
\end{pmatrix}}.
\end{equation}
\end{widetext}
Here, $k = \omega_{p} / c$ is the vacuum wavenumber of the probe field, where $c$ is the speed of light. The refraction angle $\theta_j$ at the $j$-th interface satisfies Snell's law: $n_i \sin(\theta_i) = n_j \sin(\theta_j)$, where $\theta_i$ is the incident angle. For each layer ($j=1,2,3$), $n_j$ and $d_j$ denote the refractive index and thickness, respectively. The polarization-dependent parameter is $\beta_j^s = \epsilon_j$ (TE mode) and $\beta_j^p = \mu_j$ (TM mode). The cavity's total $2 \times 2$ transfer matrix is given by
\begin{widetext}
\begin{equation}
X^{s,p}(\theta_{i},\omega_{p}) = M_1^{s,p}(\theta_{i},\omega_{p},d_1) M_2^{s,p}(\theta_{i},\omega_{p},d_2) M_1^{s,p}(\theta_{i},\omega_{p},d_1)=\begin{pmatrix}
X^{s,p}_{11} & X^{s,p}_{12} \\
X^{s,p}_{21} & X^{s,p}_{22}
\end{pmatrix},
\end{equation}
\end{widetext}
where $M_j^{s,p}(\theta_i,\omega_{p}, d_{j}$) depends on the parameters of the respective layer, and the superscript $(s, p)$ indicates the incident wave polarization.
The Fresnel reflection coefficients $r^{s}$ and $r^{p}$ are~\cite{SaeedAsiriControlling} 
\begin{equation}
r^{s,p}(\theta_{i},\omega_{p}) =\frac{\cos{\theta}(X^{s,p}_{22}-X^{s,p}_{11})-(\cos^2{\theta} X_{12}^{s,p}-X_{21}^{s,p})}{\cos{\theta}(X_{22}+X_{11})-(\cos^2{\theta} X_{12}^{s,p}+X_{21}^{s,p})}.\label{RC}
\end{equation}
Here, $X_{ij}^{s,p}$ (with $i,j=1,2$) denote the matrix elements of $X^{s,p}(\theta_{i},\omega_{p})$. Equation~(\ref{RC}) reveals that the reflection coefficients depend on the cavity permittivity $\epsilon_2$, tunable via the susceptibility $\chi$ through $\epsilon_2 = 1 + \chi$. The spin-dependent transverse shift is consequently derived from the Fresnel reflection coefficients.

We consider a Gaussian-shaped probe field incident on the interface of a cavity magnomechanical system, whose angular spectrum is given by:
\begin{equation}
\tilde{E}_i = \frac{w_{0}}{\sqrt{2\pi}} \exp\left(-\frac{w_0^2 (k_{ix}^2 + k_{iy}^2)}{4} \right),
\end{equation}
where $w_{0}$ is the beam waist radius, and 
$k_{ix}, k_{iy}$ denote the incident probe field's wavevector 
$x$- and $y$-components, respectively~\cite{Xiang2017, controllingPSHE}. Enforcement of boundary conditions~\cite{PhysRevEBliokh, Luo2011PRA, Xiang2017} yields the Fresnel coefficients, from which the reflected angular spectrum is derived,
\begin{equation}
\begin{pmatrix}
\tilde{E}^p_{r} \\
\tilde{E}^s_{r}
\end{pmatrix}
=
\begin{pmatrix}
r^{p} & \frac{k_{r y }\cot\theta_{i}(r^{p} + r^{s})}{k} \\
- \frac{k_{r y}\cot\theta_{i}(r^{p} + r^{s})}{k} & r^{s}
\end{pmatrix}
\begin{pmatrix}
\tilde{E}^p_{i} \\
\tilde{E}^s_{i}
\end{pmatrix},
\end{equation}
where $k_{ry}$ is the reflected wavevector's $y$-component. The incident angular spectra for 
$p$- and $s$-polarized probe field components are denoted by 
$\tilde{E}^p_{i}$ and $\tilde{E}^s_{i}$, respectively. We consider a 
$p$-polarized Gaussian probe field. The reflected angular spectrum determines the circular components  of the reflected field, 
$\mathbf{E}^p_{r+}$ and $\mathbf{E}^p_{r-}$, corresponding to right- ($\delta^{+}_p$) and left-handed ($\delta^{-}_p$) circularly polarized states, respectively~\cite{Xiang2017},
\begin{widetext}
\begin{equation}
\begin{aligned}
\mathbf{E}^p_{r\pm}(x_{r},y_{r},z_{r}) &= \frac{(\mathbf{e}_{rx} \pm i\mathbf{e}_{ry})}{\sqrt{\pi} w_0} \frac{z_R}{z_{R} + iz_{r}} \exp(ik_{r} z_{r}) \exp\left[ -\frac{k}{2} \frac{x_r^2 + y_r^2}{z_{R} + iz_{r}} \right] \\
&\times \left[ r^p - \frac{ix}{z_{R} + iz_{r}} \frac{\partial r^p}{\partial \theta_i} \pm \frac{y}{z_{R} + iz_r} (r^p + r^s) \pm \frac{ixy}{(z_{R} + iz_{r})^2} \left( \frac{\partial r^p}{\partial \theta_i} + \frac{\partial r^s}{\partial \theta_i} \right) \right],\label{reflected field}
\end{aligned}
\end{equation}
\end{widetext}
where $(x_r, y_r, z_r)$ denote the Cartesian coordinates of the reflected beam path, $z_{R} = k w_0^2 / 2$ is the Rayleigh length, and $\mathbf{e}_{rx}$ and $\mathbf{e}_{ry}$ are the $x$- and $y$-components, respectively, of the reflected electric field's polarization unit vector. The transverse displacements of the reflected weak probe field are then given by
\begin{equation}
\delta^{\pm}_{p} = \frac{\int y_r |E^p_{r\pm}(x_{r},y_{r},z_{r})|^2 dx_r dy_r}{\int |E^p_{r\pm}(x_{r},y_{r},z_{r})|^2 dx_r dy_r}.\label{Integralform}
\end{equation}
Based on Eqs.~(\ref{reflected field}) and (\ref{Integralform}), the spin-dependent transverse components of the reflected probe field in the effective three-layer cavity magnomechanical system can be written as~\cite{Luo2011PRA, Xiang2017} 
\begin{equation}
\delta^{\pm}_{p} = \mp \frac{k w_{0}^2 \text{Re}[1 + \frac{r^{s}}{r^{p}}]\cot\theta_{i}}{k^2w_{0}^2 +|\frac{\partial\ln r^{p} }{\partial\theta_{i}}|^2 + |(1 + \frac{r^{s}}{r^{p}})\cot \theta_{i}|^2}.\label{PSHE}
\end{equation}

\section{Results and Discussion}
\begin{figure*}
\centering
\includegraphics[width=\linewidth]{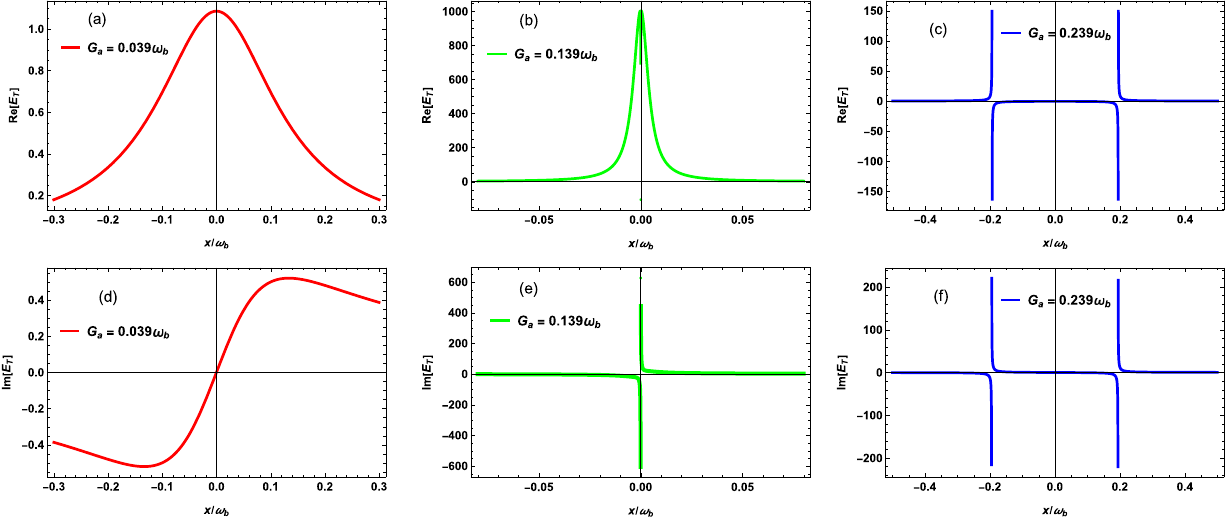}
\caption{(a-c) Absorption spectra (real part of the output field $E_T$) and (d-f) dispersion spectra (imaginary part of $E_T$) versus normalized effective detuning $x/\omega_{b}$. Columns correspond to $G_{a} = 0.039\,\omega_{b}$ (broken $\mathcal{PT}$-symmetry, red), $0.139\,\omega_{b}$ (third-order exceptional point $\mathrm{EP_3}$, green), and $0.239\,\omega_{b}$ ($\mathcal{PT}$-symmetry, blue). Fixed parameters: $\kappa_{a}/2\pi= 2.1~\mathrm{MHz}$, $\gamma_{b}/2\pi= 150~\mathrm{Hz}$, $\kappa_{m}=\kappa_{a} + \gamma_{b}$, $\omega_{b}/2\pi=15.101~\mathrm{MHz}$, and $G_{b}/2\pi=0.001~\mathrm{MHz}$.}
\label{fig3}
\end{figure*}

This section presents the numerical results. The parameters are adopted from a recent experimental study of a hybrid cavity magnomechanical system~\cite{Zhang2016cavity}: mechanical frequency $\omega_b / 2 \pi=15.101~\mathrm{MHz}$, cavity decay rate $\kappa_a / 2 \pi=2.1~\mathrm{MHz}$, cavity frequency $\omega_a / 2 \pi=1.32~\mathrm{GHz}$, mechanical damping rate $\gamma_b / 2 \pi=150~\mathrm{Hz}$, magnon gain rate $\kappa_m=\kappa_a+\gamma_b$, and effective magnomechanical coupling rate $G_b / 2 \pi=0.001~\mathrm{MHz}$. The YIG sphere has diameter $D=250~\mu \mathrm{m}$, spin density $\rho=4.22\times 10^{27}~\mathrm{m}^{-3}$, gyromagnetic ratio $\gamma / 2 \pi=28~\mathrm{GHz} / \mathrm{T}$, and is subject to a driving magnetic field $B_0 \leq 0.5~\mathrm{mT}$, corresponding to $G_b / 2 \pi \leq 1.5~\mathrm{MHz}$~\cite{Lu2021Ep}. For the PSHE investigation, the dielectric constants are $\epsilon_0=1, \epsilon_1=\epsilon_3=2.2$, with layer thicknesses $d_1=4~\mathrm{mm}$ and $d_2=45~\mathrm{mm}$~\cite{Li2020phasecontrol, Zhang2016cavity}. The incident beam is well-collimated with waist $w_0=50\,\lambda$~\cite{MuqddarPRA}. The system exhibits $\mathcal{PT}$-symmetry solely when the traveling field angle relative to the cavity $x$-axis is $\pi / 2$.

Numerical results for the PSHE are presented in three regimes: (i) broken $\mathcal{PT}$-symmetry, (ii) $\mathrm{EP_3}$, and (iii) $\mathcal{PT}$-symmetric phase, as defined in Sec.~\ref{SecIIA}. Figure~\ref{fig3} shows the probe field's absorption [$\text{Re}(E_T)$] and dispersion [$\text{Im}(E_T)$] spectra versus normalized effective detuning $x/\omega_b$ across these regimes. Panel (a) displays the broken-$\mathcal{PT}$-symmetry absorption spectrum with complex conjugate eigenvalues, exhibiting a Lorentzian peak at resonance ($x=0$). At the $\mathrm{EP_3}$ [panel (b)], eigenvalue-eigenvector coalescence induces a sharp absorption peak at $x=0$, signifying enhanced sensitivity to perturbations. The $\mathcal{PT}$-symmetric phase [panel (c)] reveals symmetric absorption resonances with balanced positive/negative amplitudes about $x=0$~\cite{Rüter2010Nature}, indicating strong magnon-photon coherence. Panels (d)-(f) present corresponding dispersion spectra. The $\mathrm{EP_3}$ [panel (e)] exhibits a divergence at $x=0$, corresponding to a rapid probe phase change. The $\mathcal{PT}$-symmetric regime [panel (f)] displays asymmetric resonances about $x=0$, characteristic of $\mathcal{PT}$-symmetric systems~\cite{Rüter2010Nature}.
\begin{figure}
\centering
\includegraphics[width=\linewidth]{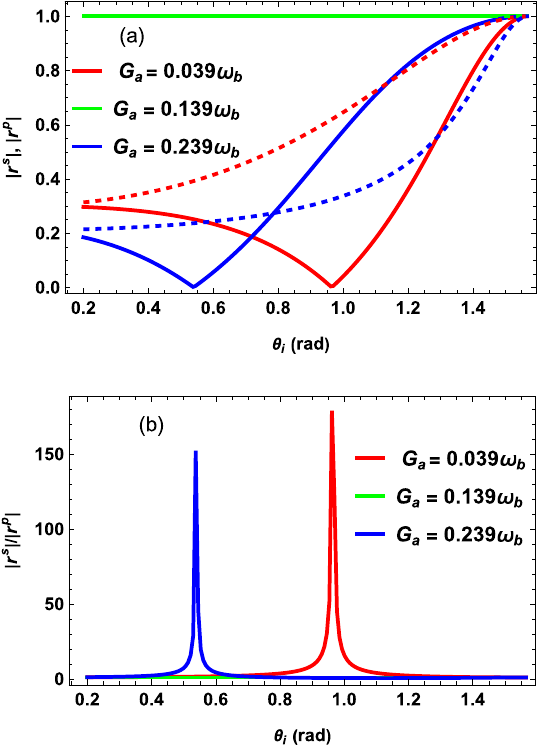}
\caption{(a) Absolute values of the reflection coefficients $|r^{s}|$ (dashed) and $|r^{p}|$ (solid), and (b) their ratio $|r^{s}|/|r^{p}|$ versus incident angle $\theta_{i}$, at $G_{a} = 0.039\,\omega_{b}$, $0.139\,\omega_{b}$, and $0.239\,\omega_{b}$, corresponding to the broken $\mathcal{PT}$-symmetric (red), $\mathrm{EP_3}$ (green), and $\mathcal{PT}$-symmetric (blue) phases at the resonance condition $x=0$, respectively. Other parameters are $\kappa_{a}/2\pi= 2.1~\mathrm{MHz}$, $\gamma_{b}/2\pi= 150~\mathrm{Hz}$, $\kappa_{m}=\kappa_{a} + \gamma_{b}$, $\omega_{b}/2\pi =15.101~\mathrm{MHz}$, $G_{b}/2\pi=0.001~\mathrm{MHz}$, $\omega_{a}/2\pi = 1.32~\mathrm{GHz}$, $w_{0}=50\,\lambda$, $d_{1} = 4~\mathrm{mm}$, $d_{2} = 45~\mathrm{mm}$, $\epsilon_{0}=1$, and $\epsilon_{1} =\epsilon_{3} = 2.2$.}
\label{fig4}
\end{figure}

Equation~(\ref{PSHE}) indicates that the spin-dependent transverse shift depends on the Fresnel reflection coefficients $|r^{s}|$ and $|r^{p}|$ for TE- and TM-polarized probe fields. We therefore analyze $|r^{s}|$ and $|r^{p}|$ versus incident angle $\theta_{i}$ (radians) in the broken $\mathcal{PT}$-symmetric phase, at the $\mathrm{EP_3}$, and in the $\mathcal{PT}$-symmetric phase. Figure~\ref{fig4}(a) displays $|r^{s}|$ and $|r^{p}|$ at resonance ($x=0$). At the $\mathrm{EP_3}$, both coefficients remain constant ($\approx 1$), indicating near-total reflection due to eigenvalue-eigenvector coalescence that suppresses spin-dependent responses. 
In broken and $\mathcal{PT}$-symmetric phases, low coefficients at small $\theta_{i}$ imply increased transmission and reduced reflection. As $\theta_{i}$ increases, $|r^{s}|$ (dashed curves) rises while $|r^{p}|$ (solid curves) decreases, vanishing at Brewster angles $\theta_{B} = 0.538$ rad ($\mathcal{PT}$-symmetric) and $\theta_{B} = 0.962$ rad (broken phase). This vanishing reflection, characteristic of TM polarization at Brewster's angle, occurs when reflected and refracted rays become orthogonal. Beyond $\theta_{B}$, $|r^{p}|$ increases, restoring TM reflection. These trends, consistent with Refs.~\cite{waseem2024Gain, MuqddarPRA}, enable transverse shift examination through reflection coefficient differences in the present non-Hermitian cavity magnomechanical system in the broken $\mathcal{PT}$-symmetric phase, $\mathrm{EP_3}$, and $\mathcal{PT}$-symmetric phase.

The ratio $|r^{s}|/|r^{p}|$ of TE- to TM-polarized reflection coefficients is analyzed next, as Eq.~(\ref{PSHE}) shows the spin-dependent transverse shift depends critically on this ratio. Enhancement occurs when $|r^{s}|/|r^{p}| > 1$. Figure~\ref{fig4}(b) plots $|r^{s}|/|r^{p}|$ versus $\theta_{i}$ at resonance ($x=0$) for broken $\mathcal{PT}$-symmetric (red), $\mathrm{EP_3}$ (green), and $\mathcal{PT}$-symmetric (blue) phases. Near Brewster angles $\theta_{B} = 0.538$ rad ($\mathcal{PT}$-symmetric) and $\theta_{B} = 0.962$ rad (broken phase), the ratio surges due to the contrasting approaches of TE- and TM-polarized components of probe field with the interface at this specific angle in $\mathcal{PT}$-symmetric and broken-$\mathcal{PT}$-symmetric phases, respectively. At these angles, $|r^{p}|$ vanishes [Fig.~\ref{fig4}(a), solid curves] from destructive interference at the interface, while $|r^{s}|$ remains finite (dashed curves). This disparity enhances $|r^{s}|/|r^{p}|$ as TM polarization refracts entirely into the second medium while TE reflects. Characterization requires narrow $\theta_{i}$ ranges near $\theta_{B}$. At the $\mathrm{EP_3}$, absent phase contrast [Fig.~\ref{fig4}(a)] maintains $|r^{s}|/|r^{p}| < 1$ [Fig.~\ref{fig4}(b)].

\begin{figure}
\centering
\includegraphics[width=\linewidth]{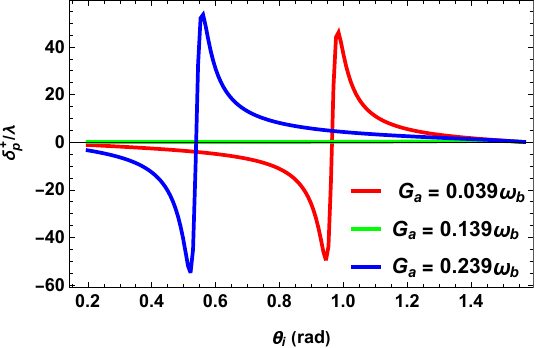}
\caption{Normalized PSHE $\delta^{+}_{p}/\lambda$ versus incident angle $\theta_{i}$ (rad) at $G_{a} = 0.039\,\omega_{b}$, $0.139\,\omega_{b}$, and $0.239\,\omega_{b}$, corresponding to the broken $\mathcal{PT}$-symmetric (red), $\mathrm{EP_3}$ (green), and $\mathcal{PT}$-symmetric (blue) phases at the resonance condition $x=0$, respectively. Other parameters are $\kappa_{a}/2\pi= 2.1~\mathrm{MHz}$, $\gamma_{b}/2\pi= 150~\mathrm{Hz}$, $\kappa_{m}=\kappa_{a} + \gamma_{b}$, $\omega_{b}/2\pi =15.101~\mathrm{MHz}$, $G_{b}/2\pi=0.001~\mathrm{MHz}$, $\omega_{a}/2\pi = 1.32~\mathrm{GHz}$, $w_{0}=50\,\lambda$, $d_{1} = 4~\mathrm{mm}$, $d_{2} = 45~\mathrm{mm}$, $\epsilon_{0}=1$, and $\epsilon_{1} = \epsilon_{3} = 2.2$.}
\label{fig5}
\end{figure}
\begin{figure*}
\begin{center}
\includegraphics[width=\linewidth]{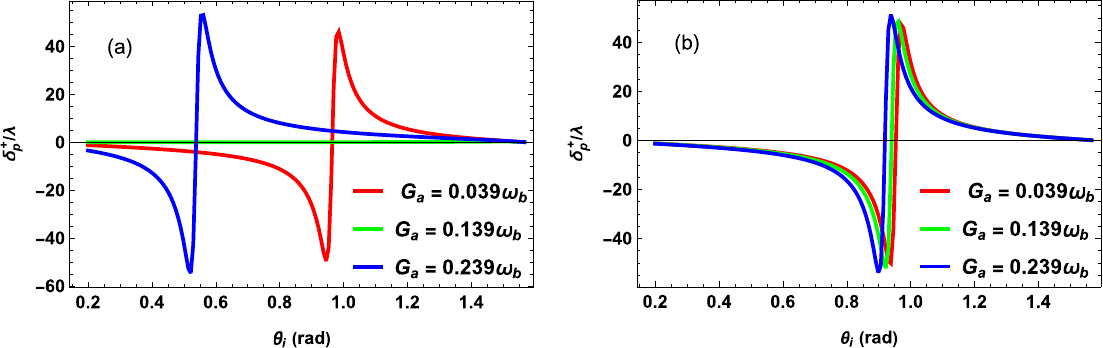}
\caption{Normalized PSHE $\delta^{+}_{p}/\lambda$ as a function of the incident angle $\theta_{i}$ (rad) at resonance ($x=0$) for effective magnon-phonon coupling strengths: (a) $G_{b}=0$ and (b) $G_{b}/2\pi=0.05~\mathrm{MHz}$. Fixed parameters: $G_{a}=0.039\,\omega_{b}$ (broken $\mathcal{PT}$-symmetric, red), $G_{a}=0.139\,\omega_{b}$ ($\mathrm{EP_2}$, green), $G_{a}=0.239\,\omega_{b}$ ($\mathcal{PT}$-symmetric, blue), $\kappa_{a}/2\pi=2.1~\mathrm{MHz}$, $\gamma_{b}/2\pi=150~\mathrm{Hz}$, $\kappa_{m}=\kappa_{a}+\gamma_{b}$, 
$\omega_{b}/2\pi=15.101~\mathrm{MHz}$, $\omega_{a}/2\pi=1.32~\mathrm{GHz}$, $w_{0}=50\,\lambda$, $d_{1}=4~\mathrm{mm}$, $d_{2}=45~\mathrm{mm}$, $\epsilon_{0}=1$, and $\epsilon_{1}=\epsilon_{3}=2.2$.}
\label{fig5ab}
\end{center}
\end{figure*}
\begin{figure*}
\begin{center}
\includegraphics[width=5.5cm]{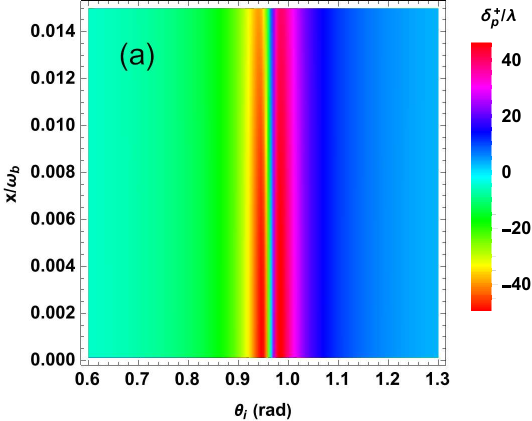}
\includegraphics[width=5.5cm]{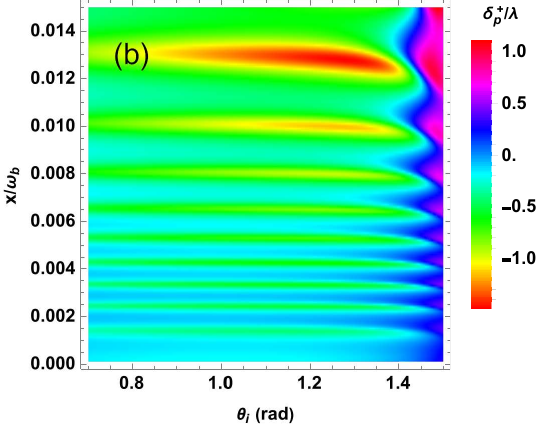}
\includegraphics[width=5.5cm]{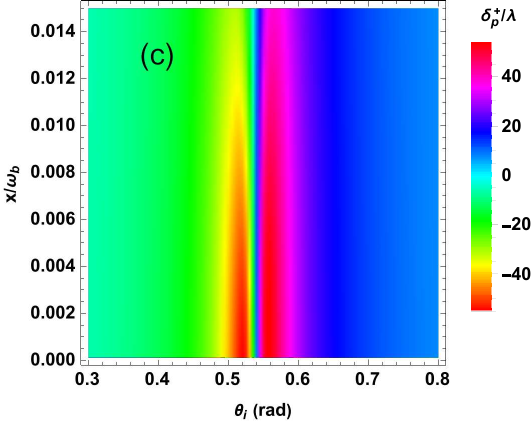}
\caption{Contour plots of normalized photonic spin Hall shift $\delta^{+}_{p}/\lambda$ versus incident angle $\theta_{i}$ and normalized effective detuning $x/\omega_{b}$ for: (a) broken $\mathcal{PT}$-symmetric phase ($G_{a} = 0.039\,\omega_{b}$), (b) $\mathrm{EP_3}$ ($G_{a} = 0.139\,\omega_{b}$), and (c) $\mathcal{PT}$-symmetric phase ($G_{a} = 0.239\,\omega_{b}$). Fixed parameters: $\kappa_{a}/2\pi = 2.1~\mathrm{MHz}$, $\gamma_{b}/2\pi = 150~\mathrm{Hz}$, $\kappa_{m} = \kappa_{a} + \gamma_{b}$, $\omega_{b}/2\pi = 15.101~\mathrm{MHz}$, $G_{b}/2\pi = 0.001~\mathrm{MHz}$, $\omega_{a}/2\pi = 1.32~\mathrm{GHz}$, $w_{0} = 50\,\lambda$, $d_{1} = 4~\mathrm{mm}$, $d_{2} = 45~\mathrm{mm}$, $\epsilon_{0} = 1$, and $\epsilon_{1} = \epsilon_{3} = 2.2$.}
\label{fig6}
\end{center}
\end{figure*}

The normalized PSHE $\delta^{+}_{p}/\lambda$, arising from the right-circularly polarized component of the reflected probe field, is analyzed. The left-circularly polarized component $\delta^{-}_{p}/\lambda$ exhibits an equal-magnitude shift in the opposite direction due to circular polarization symmetry; thus, $\delta^{+}_{p}/\lambda$ fully characterizes the transverse shift. Figure~\ref{fig5} shows $\delta^{+}_{p}/\lambda$ versus $\theta_{i}$ for broken $\mathcal{PT}$-symmetric (red), $\mathrm{EP_3}$ (green), and $\mathcal{PT}$-symmetric (blue) phases at resonance ($x=0$). 
In the $\mathcal{PT}$-symmetric phase, the shift transitions from negative ($\theta_{i} < 0.538$ rad) to positive ($\theta_{i} > 0.538$ rad). Similarly, the broken phase exhibits sign reversal at $\theta_{i} = 0.962$ rad. This sign inversion stems from a $\pi$ phase shift in reflection coefficients $|r^{s}|$ and $|r^{p}|$. Resonance enhancement ($x=0$) originates from phase differences between TE/TM Fresnel coefficients, amplifying constructive interference. The $\mathcal{PT}$-symmetric phase yields larger shifts than the broken phase, reflecting superior spin-state discrimination and coherence preservation in the non-Hermitian cavity magnomechanical system. At the $\mathrm{EP_3}$, eigenvalue-eigenvector coalescence suppresses spin-orbit interaction, inhibiting mode orthogonality and reducing the shift. These results agree with EP-mediated spin Hall control~\cite{controllingPSHE}. Such spin-orbit coupling manipulation enables spin-selective photonic devices.

Figure~\ref{fig5ab} (a) and (b) show the influence of the effective magnon-phonon coupling $G_b$ on the PSHE $\delta^{+}_{p}/\lambda$ as a function of the probe field's incident angle $\theta_i$. We compare two illustrative cases: $G_b = 0$ and $G_b = 2\pi \times0.05~\mathrm{MHz}$. For $G_b = 0$, the system exhibits a clear phase transition in the PSHE [Fig.~\ref{fig5ab}(a)], consistent with the $\mathcal{PT}$-symmetric eigenvalue spectrum [Fig.~\ref{fig2b} panel~(a)]. In contrast, when $G_b = 2\pi \times0.05~\mathrm{MHz}$, the magnon-phonon interaction not only suppresses the PSHE but also destroys the associated $\mathcal{PT}$ phase transition [Fig.~\ref{fig5ab}(b)]. This suppression is primarily attributed to enhanced absorption at resonance induced by $G_b$~\cite{MuqddarPRA}. These findings demonstrate the critical role of $G_b$ in shaping both the $\mathcal{PT}$-symmetric eigenvalue spectrum and the spin-dependent splitting of the probe field in the PSHE.

The PSHE $\delta^{+}_{p}/\lambda$ is examined versus incident angle $\theta_{i}$ and normalized effective detuning $x/\omega_{b}$ across $\mathcal{PT}$-symmetry phases. Distinct characteristics emerge in broken $\mathcal{PT}$-symmetric, $\mathrm{EP_3}$, and $\mathcal{PT}$-symmetric phases. 
The contour plots of normalized photonic spin Hall shift $\delta^{+}_{p}/\lambda$ versus incident angle $\theta_{i}$ and normalized effective detuning $x/\omega_{b}$ show that, the maximum shift transitions from negative ($\theta_{i} < 0.962$ rad) to positive ($\theta_{i} > 0.962$ rad) in the broken $\mathcal{PT}$-symmetric phase [Fig.~\ref{fig6}(a)], while suppression characterizes the $\mathrm{EP_3}$ [Fig.~\ref{fig6}(b)]. Similarly, the $\mathcal{PT}$-symmetric phase exhibits sign reversal at  $\theta_{i} = 0.538$ rad [Fig.~\ref{fig6}(c)]. This agreement validates the physical interpretation of the PSHE. 

\begin{figure*}
\begin{center}
\includegraphics[width=5.5cm]{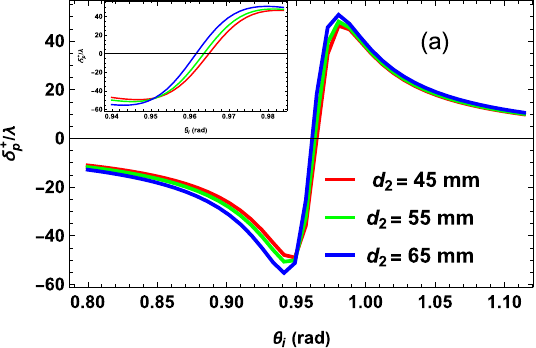}
\includegraphics[width=5.5cm]{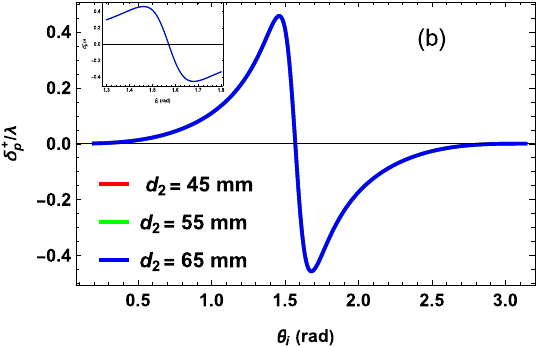}
\includegraphics[width=5.5cm]{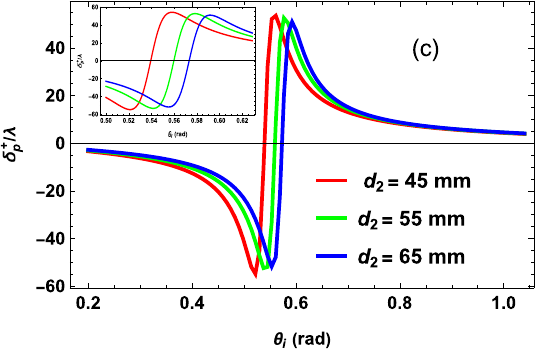}
\caption{Normalized photonic spin Hall shift $\delta^{+}_{p}/\lambda$ as a function of incident angle $\theta_{i}$ for (a) broken $\mathcal{PT}$-symmetric phase ($G_{a} = 0.039\,\omega_{b}$), (b) $\mathrm{EP_3}$ ($G_{a} = 0.139\,\omega_{b}$), and (c) $\mathcal{PT}$-symmetric phase ($G_{a} = 0.239\,\omega_{b}$), respectively at three different intracavity medium lengths. Red, green, and blue curves show the PSHE at $d_{2}=45~\mathrm{mm}$, $55~\mathrm{mm}$, and $65~\mathrm{mm}$, respectively, at the resonance condition $x=0$. The insets in Figs.~\ref{fig7}(a–c) provide magnified views of $\delta^{+}_{p}/\lambda$ as a function of $\theta_{i}$ for each phase at different intracavity medium lengths. Other parameters are $\kappa_{a}/2\pi= 2.1~\mathrm{MHz}$, $\gamma_{b}/2\pi= 150~\mathrm{Hz}$, $\kappa_{m}=\kappa_{a} + \gamma_{b}$, $\omega_{b}/2\pi =15.101~\mathrm{MHz}$, $G_{b}/2\pi=0.001~\mathrm{MHz}$, $\omega_{a}/2\pi = 1.32~\mathrm{GHz}$, $w_{0}=50\,\lambda$, $d_{1} =4~\mathrm{mm}$, $\epsilon_{0}=1$, and $\epsilon_{1} = \epsilon_{3} = 2.2$.}
\label{fig7}
\end{center}
\end{figure*}

The PSHE $\delta^{+}_{p}/\lambda$ depends critically on cavity geometry, particularly total thickness $L=2d_{1} + d_{2}$, necessitating precise dimensional control. Figures~\ref{fig7}(a-c) show $\delta^{+}_{p}/\lambda$ versus $\theta_{i}$ at resonance ($x=0$) for intracavity lengths $d_{2}$ = 45, 55, and 65~mm. 
In the broken $\mathcal{PT}$-symmetric phase [Fig.~\ref{fig7}(a)], the shift magnitude increases with $d_{2}$ while shifting toward smaller $\theta_{i}$. The $\mathrm{EP_3}$ [Fig.~\ref{fig7}(b)] exhibits length-independent suppression. The $\mathcal{PT}$-symmetric phase [Fig.~\ref{fig7}(c)] shows reduced shift magnitudes and shifts toward larger $\theta_{i}$ with increasing $d_{2}$. 
These opposing trends between symmetry phases arise from modified phase accumulation and interference conditions, demonstrating the critical role of intracavity length in PSHE control.

\section{Conclusion}
In summary, we have investigated $\mathcal{PT}$-symmetric dynamics in a hybrid non-Hermitian cavity magnomechanical system with magnon-photon and magnon-phonon interactions. Eigenvalue spectrum analysis reveals that a third-order exceptional point ($\mathrm{EP}_3$) emerges under balanced gain-loss conditions when the traveling field is oriented at $\pi/2$ relative to the cavity's $x$-axis, delineating three distinct phases: (i) broken $\mathcal{PT}$-symmetry, (ii) $\mathrm{EP_3}$, and (iii) $\mathcal{PT}$-symmetric.

The photonic spin Hall effect (PSHE) for a reflected probe field exhibits phase-dependent characteristics. At resonance ($x=0$), the $\mathcal{PT}$-symmetric phase yields significantly enhanced transverse shifts compared to the broken phase, demonstrating superior spin-state discrimination. Conversely, eigenvalue-eigenvector coalescence at the $\mathrm{EP_3}$ suppresses spin-orbit interactions, inhibiting the PSHE. Intracavity length modulation provides additional control, inducing opposing trends in shift magnitude and angular dependence between $\mathcal{PT}$-symmetric and broken phases due to modified interference conditions.
These results establish non-Hermitian cavity magnomechanics as a versatile platform for coherent spin-photon control, with promising applications in spin-selective photonic devices, quantum switching, and high-precision microwave sensing.\\

The assumption of magnon gain in our model is well-supported by established experimental techniques. In particular, gain can be realized through parametric parallel pumping, where a microwave field at twice the magnon resonance frequency $(\omega_p\approx2\omega_m)$ is applied to the magnetic medium (YIG sphere). This process parametrically amplifies the magnon population and leads to an effective negative damping (i.e., gain) of the magnon mode~\cite{BRACHER20171, Demokritov2006, Heinz2022}. In our Hamiltonian, the traveling field term represents this external pump, which can be implemented experimentally using a microwave antenna or waveguide. Recent experiments have shown that such microwave pumping not only generates pump-induced magnon modes but also drives them into strong coupling with other magnetic excitations, a phenomenon that crucially relies on the presence of magnonic gain~\cite{PRL2023, Zhijian2024}. These demonstrations confirm the experimental feasibility and practical relevance of the gain mechanism incorporated in our model.

\section*{Acknowledgement}
We acknowledge the financial support from the NSFC under grant 
No. 12174346.\\ 

\section*{References}

\bibliography{refr}

\end{document}